\documentclass[aps,prb,reprint,superscriptaddress,amsmath,amssymb,showpacs]{revtex4-2}%

\renewcommand{\vec}[1]{\mathbf{#1}}

\usepackage{braket}
\usepackage{subfigure}
\usepackage{graphicx}
\usepackage{xcolor}
\usepackage{dsfont}
\usepackage{pstricks}
\usepackage{mathrsfs}
\usepackage{bbm}
\usepackage{nicefrac}
\usepackage{mathtools}

\begin{document}

\title{Quantifying the influence of the initial state on the dynamics of an open quantum system}

\author{S.\ Wenderoth}
\affiliation{Institute of Physics, University of Freiburg, Hermann-Herder-Str. 3, D-79104 Freiburg, Germany
}

\author{H.-P. Breuer}
\affiliation{Institute of Physics, University of Freiburg, Hermann-Herder-Str. 3, D-79104 Freiburg, Germany
}
\affiliation{EUCOR Centre for Quantum Science and Quantum Computing, University of Freiburg, Hermann-Herder-Str. 3, D-79104 Freiburg, Germany
}

\author{M.\ Thoss}
\affiliation{Institute of Physics, University of Freiburg, Hermann-Herder-Str. 3, D-79104 Freiburg, Germany
}
\affiliation{EUCOR Centre for Quantum Science and Quantum Computing, University of Freiburg, Hermann-Herder-Str. 3, D-79104 Freiburg, Germany
}

\date{\today}

\begin{abstract}
A small system in contact with a macroscopic environment usually approaches an asymptotic state, determined only by some macroscopic properties of the environment such as the temperature or the chemical potential. In the long-time limit, the state of the small system is thus expected to be independent of its initial state. In some situations, however, the asymptotic state of the system is influenced by its initial state and some information about the initial state is kept for all times. Motivated by this  finding, we propose a measure to quantify the influence of the initial state of an open system on its dynamics. Using this measure we derive conditions under which the asymptotic state exists and is unique. We demonstrate our concepts for the dynamics of the spin-boson model, identify three qualitatively different long-time behaviors, and discuss how they can be distinguished based on the proposed measure.
\end{abstract}

\maketitle

\section{Introduction}
If a small system is put into contact with a macroscopic environment, the system comes to an equilibrium. This phenomenon, usually called thermalization, is known for centuries. However, its emergence from the microscopic dynamics, described by Schrödinger's equation, is still not fully understood. To make this problem more accessible, thermalization is decomposed into different aspects, which can be discussed separately. Following this approach \cite{Linden2009, Gogolin2016}, a system interacting with some environment is said to equilibrate if its state evolves towards some particular state, called the equilibrium or asymptotic state, and remains close to it for almost all times. The system thermalizes if the asymptotic state fulfills the following three properties. First, the asymptotic state is influenced by the initial state of the environment only via some macroscopic properties, like the total energy. Second, it is independent of the initial state of the small system. Third, it is close to a Gibbs or thermal state.

Recently, it was shown by Linden and coworkers \cite{Linden2009} that the equilibration of small subsystems is a general property of quantum many-body systems. Assuming that the time-evolution of the joint system, i.e. system and environment, is unitary, they showed that any small subsystem approaches an asymptotic state and remains close to it for almost all times, provided that the total Hamiltonian has non-degenerate energy level spacing \cite{Linden2009}. They also showed that the asymptotic state does not depend on the precise initial state of the environment, and analyzed in which situations the asymptotic state is independent of the initial state of the small system.

The discussion in Ref.\,[\onlinecite{Linden2009}] demonstrates that subsystem equilibration can, in fact, be derived from the Schrödinger equation if one assumes certain spectral properties of the underlying Hamiltonian. On the other hand, it is known that there are systems in contact with an environment which do not equilibrate or thermalize \cite{Chakravarty1982,Bray1982,Lidar1998,Apollaro2011,Wilner2013}. Subsystems of many-body localized systems also fail to thermalize locally due to the absence of transport in these systems \cite{Anderson1958, Basko2006, Nandkishore2015}. 

For finite dimensional systems the spectrum of the total Hamiltonian can be used to investigate subsystem equilibration. There are, however, some limitations to this approach. First, the spectrum can only be calculated for relatively small systems. Second, even if it was possible to calculate the spectrum for arbitrarily large systems, this approach cannot be used to investigate subsystem equilibration for infinite dimensional systems, which occur for example in the description of condensed-phase environments, as the spectrum of such systems is typically continuous. Third, non-degenerate energy level spacing of the Hamiltonian only guarantee two aspects of thermalization: subsystem state equilibration and bath state independence.

The dependence of the equilibrium state on the initial state of the subsystem has been investigated in \cite{Smirne2012}, employing the time average of the quantum dynamical map which encodes the time evolution of a subsystem in contact with some environment. The analysis of \cite{Smirne2012} focuses on the role of system-environment correlations and changes in the environmental states in the thermalization of open quantum systems. Here, we study the full time dependence of the quantum dynamical map. This not only allows us to discuss the influence of the initial state on the equilibrium state, but also to derive an upper bound for the impact of the initial state on expectation values at all times. Compared to the analysis based on spectral properties of the total Hamiltonian, an analysis of the dynamical map is much more feasible, as its dimension is determined by the subsystem.

A simple example for a dynamical map is provided by a dynamical semigroup with a time-independent generator in Lindblad form \cite{Lindblad1976, Gorini1976}. For this special case some properties of the Lindblad generator which guarantee a unique asymptotic state are known. These include certain properties of the Kossakowski matrix \cite{Spohn1976} or the algebra of the Lindblad generators \cite{Spohn1977, Frigerio1977, Frigerio1978, Evans1977}. On the other hand, it is known that even for the simple situation of a time-independent generator in Lindblad form, the asymptotic state can exhibit a dependence on the initial state of the open system if the generator possesses symmetries \cite{Dietz2003,Baumgartner2008a,Baumgartner2008b,Buca2012,Albert2014}.

In general, the dynamics of an open system are, however, not described by a dynamical semigroup with a generator in Lindblad form. In such situations, the classification of the asymptotic behavior of the open system cannot be based on properties of the generator. In this work, without assuming a particular form of the dynamical map, we propose a measure to quantify the influence of the initial state of an open system on its dynamics. This measure can be used to analyze if an open system approaches an asymptotic state and can be used to quantify the influence of the initial state on the asymptotic state.

The paper is organized as follows: In Sec.\,\ref{sec:Theory} we briefly review some basic concepts of the theory of open quantum systems, introduce the dynamical map and discuss how it can be used to quantify the influence of the initial state on the dynamics of an open system. To illustrate our theoretical concepts we demonstrate them for a two-level system coupled to a harmonic environment, also known as the spin-boson model. We introduce the model as well as the method we use to simulate the dynamics, the multilayer multiconfiguration time-dependent Hartree approach (ML-MCTDH), in Sec.\,\ref{sec:Model}. In Sec.\,\ref{sec:Results} we apply our measure to the spin-boson model and discuss different long-time behaviors.

\section{Theory}
\label{sec:Theory}
An open quantum system $S$ can be considered as a subsystem of a larger system, composed of $S$ and another subsystem $E$, its environment \cite{Alicki2007, Breuer2007, Davies1976}. The Hilbert space of the joint system $S+E$ is given by
\begin{align}
\mathcal{H}_{SE} &= \mathcal{H}_S \otimes \mathcal{H}_E,
\end{align}
where $\mathcal{H}_S$ and $\mathcal{H}_E$ denote the Hilbert spaces of $S$ and $E$, respectively. Physical states of the joint system are represented by linear, self-adjoint, positive semidefinite operators of unit trace on $\mathcal{H}_{SE}$, also called density matrices. We denote the set of all physical states over a Hilbert space $\mathcal{H}$ by $\mathcal{S}(\mathcal{H})$. The state of the open system $S$ is obtained by tracing out the environmental degrees of freedom of $\rho_{SE}\in \mathcal{S}(\mathcal{H}_{SE})$, i.e. $\rho_S = {\rm tr}_E \{\rho_{SE}\}$, where ${\rm tr}_E \{\cdot\}$ denotes the partial trace over the environment. Throughout this paper we consider finite dimensional open systems $S$, and thus, the state of the open system can be represented as a finite, Hermitian, positive semidefinite matrix with trace one.

We suppose that the joint system is closed and described by a time-independent Hamiltonian of the form
\begin{align}
H = H_S + H_E + H_{I},
\end{align}
where $H_S$ ($H_E$) describes the Hamiltonian of the system (environment), respectively, and $H_{I}$ describes the interaction between the system and the environment.

Since the joint system is closed, its time-evolution is described by a unitary time evolution operator $U(t)={\rm e}^{-i H t/\hbar}$. In the following we set $\hbar=1$. Under the assumption of a factorized initial state, i.e.  $\rho_{SE}(0) = \rho_S(0) \otimes \rho_E(0)$, the state of the open system $\rho_S(t)$ at time $t$ is given by
\begin{align}
\rho_S(t) &= {\rm tr}_E \{ {\rm e}^{-i H t} \rho_S(0) \otimes \rho_E(0) {\rm e}^{i H t}\} \nonumber \\
 & = \Phi_t \rho_S(0).\label{eq:OpenSystemTimeEvolution}
\end{align}

For a fixed initial state of the environment $\rho_E(0)$, Eq.\,(\ref{eq:OpenSystemTimeEvolution}) defines a linear map $\Phi_t$ on the set $\mathcal{S}(\mathcal{H}_S)$ as
\begin{align}
\Phi_t : \mathcal{S}(\mathcal{H}_S) &\rightarrow \mathcal{S}(\mathcal{H}_S) \nonumber \\
\rho_S(0) &\mapsto \rho_S(t).\label{eq:DefinitionDynamicalMap}
\end{align}
The map $\Phi_t$, called the dynamical map, is a superoperator mapping any initial state of the open system to the corresponding state at time $t$. Thus, it encodes the complete information on the time evolution of the open system. Using Eq.\,(\ref{eq:OpenSystemTimeEvolution}), one can show that the dynamical map preserves the Hermiticity and the trace of operators, and that it is a positive map, i.e. $\Phi_t$ maps positive operators to positive operators. Hence, $\Phi_t$ maps physical states to physical states, implying that $\mathcal{S}(\mathcal{H}_S)$ is closed under the action of $\Phi_t$. Note that the dynamical map is not only positive but also completely positive \cite{Alicki2007, Breuer2007, Davies1976}.

To discuss the long-time behavior of the open quantum system, we introduce the notion of an asymptotic state $\rho_{S,\infty}$. For a given initial state $\rho_S(0)$, the corresponding asymptotic state is defined as
\begin{align}
\rho_{S, \infty} \coloneqq \lim_{t \to \infty} \Phi_t \rho_S(0),\label{eq:DefAsymptoticState}
\end{align}
provided the limit exists. In general, the asymptotic state can depend on the initial state. The set of all asymptotic states is given by the image of all possible initial states of the open system under the dynamical map. Formally it is defined as $\textrm{Im} \Phi_{\infty} \coloneqq \lim_{t \to \infty} \Phi_t (\mathcal{S}(\mathcal{H}_S))$. A state $\rho_S$ is said to be invariant if it is left unchanged by the dynamical map, i.e. $ \rho_S = \Phi_t \rho_S$ holds for all times $t$. We emphasize that an asymptotic state need not be an invariant state, i.e. $\Phi_t\rho_{S,\infty} \neq \rho_{S,\infty}$. The dynamical map $\Phi_t$ is called relaxing if there exists a unique state $\tilde{\rho}_{\infty} \in \mathcal{S}(\mathcal{H}_S)$, such that
\begin{align}
\tilde{\rho}_{\infty} = \lim_{t \to \infty} \Phi_t \rho_S(0)
\end{align}
holds for all possible initial states $\rho_S(0) \in \mathcal{S}(\mathcal{H}_S)$ \cite{Spohn1976, Spohn1977}. In such a situation, the set of asymptotic states $\textrm{Im} \Phi_{\infty}$ consists of a single state, i.e. is zero dimensional. There are situations, in which $\rho_S(t)$ does not become stationary as $t \to \infty$, and the limit in Eq.\,(\ref{eq:DefAsymptoticState}) does not exist. This happens, for example, if decoherence free subspaces in the Hilbert space of the open system exist \cite{Lidar1998}.

To investigate the asymptotic behavior of the open system, we consider the dynamical map $\Phi_t$, defined by Eqs.\,(\ref{eq:OpenSystemTimeEvolution}) and (\ref{eq:DefinitionDynamicalMap}), which encodes the full information about the dynamics of the open system. Calculating the dynamical map is usually not possible since it involves the time-evolution of the joint system. It is, however, possible to reconstruct the dynamical map from the dynamics of the open system alone in the following way. Let $\rho_{S,kl}$ be the matrix representation of the reduced density matrix, i.e. $\rho_{S,kl}=\braket{k|\rho_{S}|l}$, where $\ket{l}$ is some fixed basis in the Hilbert space of the open system. The action of the dynamical map on the reduced density matrix can be written as
\begin{align}
\rho_{S,ij}(t) &= \sum_{kl} \Phi_{t;ij,kl} \rho_{S,kl}(0), \label{eq:RhoReducedSystemPropagator}
\end{align}
where $\Phi_{t;ij,kl} = {\rm tr}\big\{(\ket{i}\bra{j})^\dagger \Phi_t \ket{k} \bra{l} \big\}$. In this representation the dynamical map is a rank-4 tensor with $N^4$ complex elements. The time evolution of $N^2$ different initial states can be used to invert Eq.\,(\ref{eq:RhoReducedSystemPropagator}) thus allowing to reconstruct the dynamical map from the dynamics of the open system alone. This approach was, for example, used by Kidon et al.\,to obtain the exact memory kernel for the generalized non-equilibrium Anderson-Holstein model \cite{Kidon2015, Kidon2018}. Any suitable impurity solver can be used to calculate the time-evolution of the open system \cite{Cohen2013,Wang2009,Cohen2015,Tanimura2020}. To obtain the dynamical map numerically we use Eq.\,(\ref{eq:RhoReducedSystemPropagator}), which defines the dynamical map as a map acting on a matrix representation of the reduced state of the open system. To analyze the influence of the initial state on its dynamics it is more convenient to use the representation of the state of the open system in terms of the Bloch vector.

The density matrix of the open system can be represented by the $(N^2-1)$ dimensional Bloch vector \cite{Bloch1946, Hioe1981} or coherence vector \cite{Lendi1986,Alicki2007}. This means that the state $\rho_S$ is expanded in terms of the $(N^2-1)$ Hermitian and traceless generators of ${\rm SU}(N)$ as
\begin{align}
\rho_S = \frac{1}{N}\mathds{1} + \frac{1}{2} \sum_{n=1}^{N^2-1} a_n {\bf T}_n,\label{eq:RhoBlochExpansion}
\end{align}
where $\mathds{1}$ is the identity matrix, the matrices $\{{\bf T}_n\}$ are the $(N^2-1)$ generators of ${\rm SU}(N)$, and $\{ a_n\}$ constitutes the $(N^2-1)$ dimensional Bloch vector, i.e. $a_n = {\rm tr} \{ \rho T_n \}$. For a definition of the matrices ${\bf T}_n$ see, for example, \cite{Kimura2003, Alicki2007}. Thus, every state $\rho_S$ is represented by a unique element of $\mathds{R}^{N^2-1}$. This representation guarantees hermiticity and unity of the trace but not positivity. Thus, not all elements of $\mathds{R}^{N^2-1}$ represent physical states. The set of physical states is only a subset of $\mathds{R}^{N^2-1}$, denoted by $B(\mathds{R}^{N^2-1})$ which represents $\mathcal{S}(\mathcal{H}_S)$ and is sometimes called the Bloch-vector space \cite{Kimura2003, Kimura2005}. For $N=2$, $B(\mathds{R}^{N^2-1})$ is the well-known Bloch ball. For $N\geq3$ only some general properties of $B(\mathds{R}^{N^2-1})$ were proven \cite{Jakobczyk2001, Kimura2003, Byrd2003, Kimura2005}. For our discussion, however, it is sufficient to know that $B(\mathds{R}^{N^2-1})$ is mapped into itself under the dynamical map, which is guaranteed by the definition of the dynamical map.

To obtain the action of the dynamical map on the Bloch vector one can employ the fact that the dynamical map is completely positive and trace preserving, and thus, can be represented in terms of a set of Kraus operators $B_n$ as \cite{Kraus1983, Alicki2007}
\begin{align}
\rho_S(t) &= \sum_{n=1}^{N^2} B_n(t) \rho_S(0) B_n^\dagger(t),
\end{align}
with $\sum_{n=1}^{N^2}B_n^\dagger(t) B_n(t) = \mathds{1}$. Using this representation and the expansion of the reduced density matrix in terms of the generalized Bloch vector given by Eq.\,(\ref{eq:RhoBlochExpansion}) the action of the dynamical map on the generalized Bloch vector, denoted by $\phi_t$,  can be written as \cite{Nielsen2000, Alicki2007}
\begin{align}
\phi_t : B(\mathds{R}^{N^2-1}) &\rightarrow B(\mathds{R}^{N^2-1}) \nonumber \\
\vec{a}(0) &\mapsto \vec{a}(t) = {\bf M}(t)\vec{a}(0)+  \vec{b}(t).\label{eq:BlochVectorTimeEvolution}
\end{align}
Here $\vec{b}(t)\in \mathds{R}^{N^2-1}$ and ${\bf M}(t)\in \mathds{R}^{(N^2-1) \times (N^2-1)}$. The map $\phi_t$ defines an affine transformation on $B(\mathds{R}^{N^2-1})$ relating the initial Bloch vector $\vec{a}(0)$ to the corresponding Bloch vector $\vec{a}(t)$ at time $t$.

Equation (\ref{eq:BlochVectorTimeEvolution}) is the starting point of our analysis. First note that if the time-evolution of the open system is unitary, e.g. for vanishing system-environment coupling, one can show that $\vec{b}(t)=0$ and ${\bf M}^T(t){\bf M}(t)=\mathds{1}$, i.e. ${\bf M}(t)$ is an orthogonal matrix. The first equation can be shown by considering the action of the dynamical map on the vector $\mathbf{0} =(0~0~...)^T$. From Eq.\,(\ref{eq:RhoBlochExpansion}) it follows that $\mathds{1}$ is conserved under a unitary transformation. Thus, $\phi_t \mathbf{0} = \mathbf{0}$, which directly implies $\mathbf{b}(t)=0$ for all times. The orthogonality of $\mathbf{M}(t)$ follows from $\mathbf{b}(t)=\mathbf{0}$ and the fact that the Euclidean norm of the Bloch vector is preserved under unitary time evolution \cite{Alicki2007}. This means that for a unitary time-evolution $||\vec{a}(t)||_2 = ||\vec{a}(0)||_2$ holds for all times. The Euclidean norm $||\vec{x}||_2$ of a vector $\vec{x}$ is defined as $||\vec{x}||_2 = \sqrt{\sum_n x_n^2} $.

For a general $\phi_t$, however, ${\bf M}(t)$ need not to be orthogonal, and thus, is not necessarily diagonalizable. To analyze the asymptotic state of the open system we thus make use of the singular value decomposition given by
\begin{align}
{\bf M}(t) &= {\bf V}(t) {\bf S}(t) {\bf W}^T(t).
\end{align} 
Since ${\bf M}(t)$ has real entries, ${\bf V}(t)$ and ${\bf W}(t)$ can be chosen to be real, orthogonal matrices and ${\bf S}(t)$ is a positive-semidefinite diagonal matrix.

We start our analysis of the influence of the initial state by considering its influence on the expectation value of an observable $O$. To quantify this influence, we consider two different initial states of the open system, $\rho_S^1(0)$ and $\rho_S^2(0)$, and define the quantity
\begin{align}
\delta_{1,2}(t;O) &= | {\rm tr} \big\{O\big(\rho_S^1(t)-\rho_S^2(t)\big) \big\}|.
\end{align}
$\delta_{1,2}(t;O)$ describes the difference of the expectation value of $O$ at time $t$ between the two different initial states of the open system. One can show that this quantity is bounded by
\begin{align}
\delta_{1,2}(t;O) &\leq \frac{N^{3/2}}{\sqrt{2}} |o_{\rm max}| ~  S_{\rm max}(t)~ ||\vec{a}^1(0)-\vec{a}^2(0)||_2. \label{eq:BoundExpectationValues}
\end{align}
Here $o_{\rm max}$ is the eigenvalue of $O$ with the largest absolute value, $S_{\rm max}(t)$ is the largest singular value of ${\bf M}(t)$ at time $t$ and $||\vec{x}||_2$ is the Euclidean norm of the vector $\vec{x}$. The proof of Eq.\,(\ref{eq:BoundExpectationValues}) is provided in Appendix \ref{appendix:A}. Since Eq.\,(\ref{eq:BoundExpectationValues}) holds for any observable $O$ we conclude that the largest singular value of ${\bf M}(t)$ is a measure for the influence of the initial state on the state at time $t$. Consequently, the influence of the initial state on the asymptotic state can be quantified by 
\begin{align}
S_{\textrm{max}, \infty } \coloneqq \lim_{t\to \infty} S_{\rm max}(t),
\end{align}
provided that the limit exists. We note that the bound for $\delta_{1,2}(t;O)$ is very general. We will demonstrate later that for a specific observable one can find tighter bounds by employing properties of the observable of interest and their relation to the generators of ${\rm SU}(N)$.

Next, we discuss the existence and uniqueness of an asymptotic state of the open system. From Eq.\,(\ref{eq:BlochVectorTimeEvolution}) it directly follows that if the two limits
\begin{align}
{\bf b}_\infty &= \lim_{t\to \infty} \vec{b}(t), \label{eq:LimitB} \\
{\bf M}_\infty &= \lim_{t\to \infty} {\bf M}(t) \label{eq:LimitM}
\end{align}
exist, every initial state has an asymptotic state, which is given by
\begin{align}
\vec{a}_{\infty} &= \vec{b}_\infty + {\bf M}_{\infty} \vec{a}(0).\label{eq:AsymptoticState}
\end{align}
In general, the initial state $\vec{a}(0)$ has an influence on the asymptotic state. We see from Eq.\,(\ref{eq:AsymptoticState}) that the asymptotic state becomes independent of the initial state if all possible initial states are mapped to the same vector $\vec{b}_{\infty} \in \mathds{R}^{N^2-1}$. This is exactly the case if the image of ${\bf M}_\infty$ is zero dimensional, i.e. if ${\bf M}_{\infty} = \mathbf{0}$. This follows from the fact that for any dimension of the open system the Bloch-vector space $B(\mathds{R}^{N^2-1})$ includes the $(N^2-1)$ dimensional sphere with radius $r_s = \sqrt{\frac{2}{N(N-1)}}$ \cite{Kossakowski2003}. The image of this sphere under $\phi_t$ is zero dimensional if and only if $\mathbf{M}_{\infty} = \mathbf{0}$. In this case, Eq.\,(\ref{eq:AsymptoticState}) becomes independent of the initial state and the unique asymptotic state is given by $\vec{b}_{\infty}$. Note that in this case $\delta_{1,2}(t;O) \to 0$ as $t \to \infty$ independently of $O$, implying that the expectation value of any observable becomes independent of the initial state $\vec{a}(0)$. 

We conclude that the asymptotic state exists for all initial states if the limits (\ref{eq:LimitB}) and (\ref{eq:LimitM}) exist. The dynamical map is relaxing, i.e. the asymptotic state is unique, if and only if all singular values of ${\bf M}(t)$ decay to zero as $t \to \infty$. If the two limits (\ref{eq:LimitB}) and (\ref{eq:LimitM}) exist, but $\mathbf{M}_\infty \neq \mathbf{0}$ then the asymptotic state is not unique. If, on the other hand, one of the quantities $\vec{b}(t)$ or ${\bf M}(t)$ remain time-dependent at all times, there is at least one initial state for which the asymptotic state does not exist. Note that it is possible that the asymptotic state exists for some initial states, whereas for others the state of the open system remains time-dependent at all times. In such a situation, some singular values of ${\bf M}(t)$ become stationary and others remain time dependent.

\section{Model and Method}
\label{sec:Model}
To illustrate the theoretical results obtained in Sec.\,\ref{sec:Theory}, we consider the spin-boson model. The spin-boson model involves a spin, or more generally a two-level system, interacting linearly with a bath of harmonic oscillators \cite{Weiss1999, Leggett1987}. Despite its simple form, the spin-boson model exhibits several interesting effects, such as a transition from coherent dynamics to incoherent decay and a quantum phase transition \cite{Bray1982,Chakravarty1982,Wang2019}, and has been used to describe a variety of different processes and phenomena, including electron transfer \cite{Marcus1985} and macroscopic quantum coherence \cite{Weiss1987}. As we will demonstrate later, the spin-boson model also exhibits three qualitatively different long-time behaviors making this model a well-suited prototype to demonstrate and discuss the above introduced concepts. 

For the purpose of this paper it is sufficient to consider the unbiased spin-boson model. Employing mass-weighted coordinates, the Hamiltonian reads
\begin{align}
H &= \Delta \sigma_x + \frac{1}{2}\sum_{n=1}^N (p_n^2 + \omega_n^2 q_n^2) + \sigma_z \sum_{n=1}^N c_n q_n, \label{eq:HamiltonianSBM}
\end{align}
where $\sigma_x$ and $\sigma_z$ are the Pauli matrices, $\Delta$ denotes the tunneling between the two spin states, and $\omega_n$, $q_n$, and $p_n$ represent the frequency, position, and momentum of the bath oscillators, respectively; $c_n$ denotes the coupling strength of the spin to the $n$-th harmonic oscillator of the bath. The properties of the bath which influence the spin are summarized by the spectral density \cite{Weiss1999, Leggett1987} 
\begin{align}
J(\omega) &= \frac{\pi}{2} \sum_{n=1}^N \frac{c_n^2}{\omega_n} \delta(\omega-\omega_n).\label{eq:SpectralDensity}
\end{align}

To realize different long-time behaviors of the spin we consider two different functional forms of the spectral density. The first is the well-known Ohmic spectral density defined as \cite{Weiss1999, Leggett1987}
\begin{align}
J_{O}(\omega) &= \frac{\pi}{2} \alpha \omega {\rm e}^{-\omega/\omega_c},\label{eq:SpectralDensityOhmic}
\end{align}
where $\alpha$ denotes the system-bath coupling strength and $\omega_c$ the characteristic frequency of the bath. For this spectral density, it is known that in the scaling regime, $\omega_c \gg \Delta$, the spin relaxes to a unique asymptotic state for $\alpha<1$, whereas the spin localizes for $\alpha>1$ and $T=0$ \cite{Leggett1987,Weiss1999,Bray1982,Chakravarty1982,Spohn1989}.

\begin{figure}[t]
\includegraphics[scale=0.5]{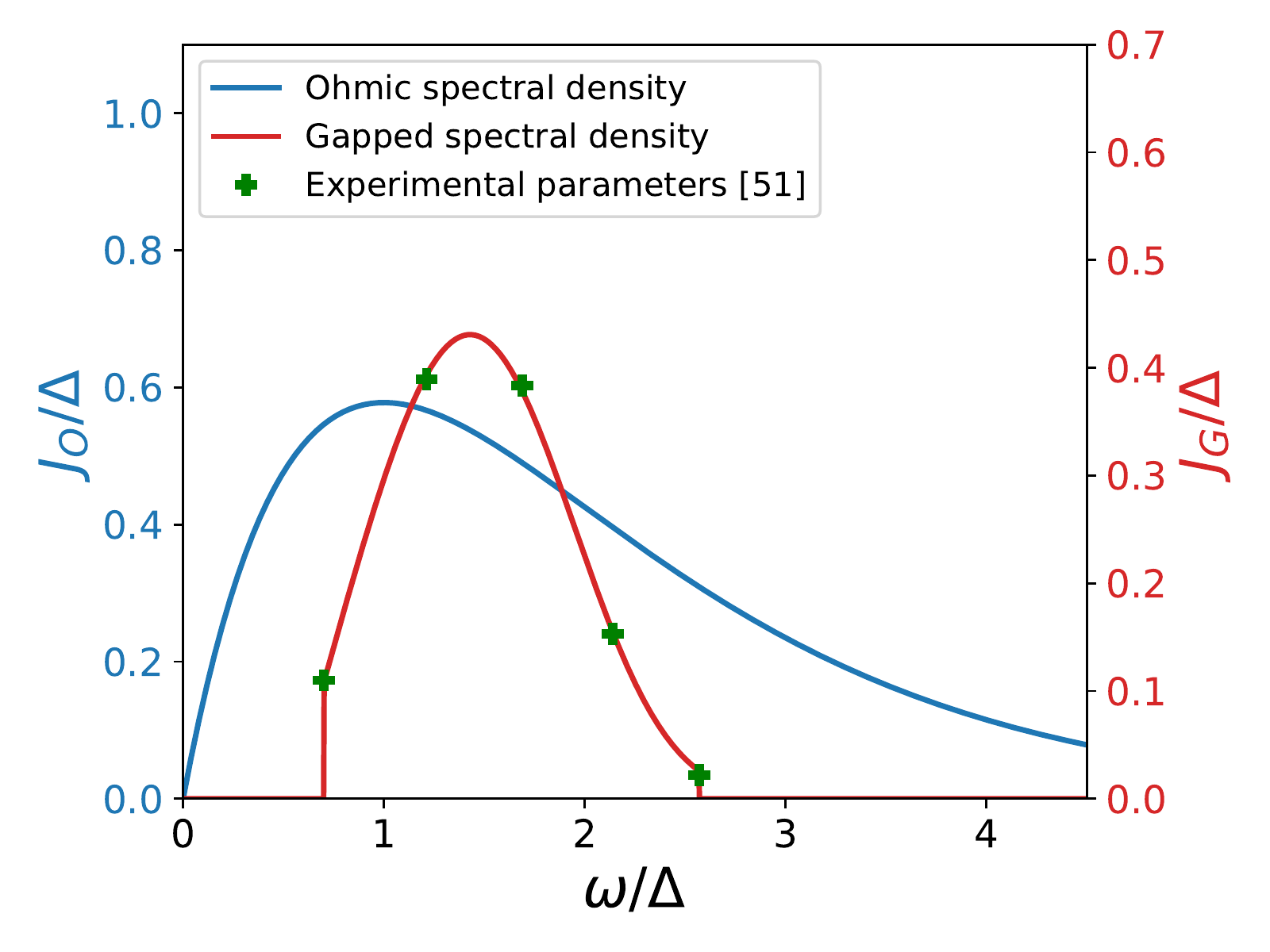}
\caption{Spectral densities considered in the model for $\alpha=1$. The blue line (left axes) shows the Ohmic spectral density for $\omega_c=\Delta$. The red line (right axes) shows the gapped spectral density. The green markers represent the experimental parameters \cite{Clos2016}. Note that the gapped spectral density is zero below $\omega_{\rm min}$ and above $\omega_{\rm max}$.}\label{fig:spectral_density}
\end{figure}

The second model we consider is inspired by a recent experimental realization of the spin-boson model using trapped ions \cite{Clos2016}. The continuum limit of the spectral density was obtained by fitting a continuous function to the parameters of an experimental realization of the spin-boson model with 5 environmental modes. The distinct feature of this model is that the spectral density is gapped, i.e. the spectral density is zero below some $\omega_{\min}>0$. This results in a spectral density of the form 
\begin{align}
J_{G}(\omega) &= \frac{\pi}{2} \alpha a(\omega - b) {\rm e}^{-(\frac{\omega-b}{c})^3} \chi_{[\omega_{\rm min}, \omega_{\rm max}]},\label{eq:SpectralDensityGapped}
\end{align}
where $\alpha$ denotes the coupling strength, $\chi_{[x, y]}$ denotes the characteristic function of the interval $[x, y]$, and $a$, $b$, and $c$ are fitting parameters. The spectral density $J_G(\omega)$ has a maximum at the transition frequency of the spin, i.e. at $\omega=2\Delta$. A comparison between the two spectral densities showing the qualitative differences is given in Fig.\,\ref{fig:spectral_density}. A more detailed description of the origin of the model and some details about the fit are provided in Appendix \ref{appendix:B}.

To simulate the dynamics of the spin-boson model, we employ the multilayer multi-configuration time-dependent Hartree (ML-MCTDH) approach \cite{Wang2003, Manthe2008, Vendrell2011, Wang2015}, which allows us to propagate the wave function of the joint system in a numerically exact way. The ML-MCTDH approach represents a rigorous variational basis-set method, which uses a multi-configuration expansion of the wave function $\ket{\Psi(t)}$, employing time-dependent basis functions and a hierarchical multilayer representation. Specifically, a representation of the wave function $\ket{\Psi(t)}$ which corresponds to a hierarchical tensor decomposition in the form of a tensor tree network is employed. Within this approach, the wave function is recursively expanded as a superposition of Hartree products, the so-called "single-particle functions" (SPFs). The hierarchy is terminated by expanding the SPFs in the deepest layer in terms of time-independent basis functions/configurations, each of which may contain several physical degrees of freedom. For more technical details, we refer the reader to earlier work on the ML-MCTDH approach and its applications to the spin-boson model \cite{Wang2003, Manthe2008, Vendrell2011, Wang2015, Wang2008, Wang2019}. The ML-MCTDH equations of motion for the expansion coefficients and the SPFs are obtained by applying the Dirac-Frenkel variational principle \cite{Wang2003,Wang2009}, thus ensuring convergence to the solution of the time-dependent Schrödinger equation upon increasing the number of variational parameters included in the calculation. 

The ML-MCTDH approach allows for the simulation of large but finite quantum systems. Thus, we represent the continuous bath by a finite number of modes. In this work we use an equidistant distribution but other choices are possible \cite{Wang2008, Vega2015}. To ensure convergence to the continuum limit over the timescale considered, we employ several hundreds of modes. For a detailed discussion of the numerical treatment of a continuous bath, see Ref.\,[\onlinecite{Wang2008}].

Here, we employ an implementation of the ML-MCTDH theory with up to four dynamical layers plus one static layer. To ensure that convergence is achieved, for each set of physical parameter a series of careful convergence tests were performed with respect to all the variational parameters such as the number of bath modes, primitive basis functions and SPFs in each layer.

\section{Results and discussion}
\label{sec:Results}
In this section, we use the theoretical concepts introduced above to analyze the initial state dependence for the example of the spin-boson model. In the limit $t \to \infty$, an open system can exhibit three qualitatively different asymptotic behaviors. First, the open system can relax to a unique asymptotic state. Second, the open system can relax to an asymptotic state which depends on the initial state. Last, the open system may not relax to an asymptotic state and $\rho_S(t)$ remains time-dependent at all times. As we will demonstrate below, all these cases can be distinguished by properties of the matrix ${\bf M}(t)$ introduced in Sec.\,\ref{sec:Theory}. 

To obtain ${\bf M}(t)$ and $\vec{b}(t)$ numerically we proceed as follows. Employing the ML-MCTDH method we simulate the dynamics for four linearly independent initial states of the reduced density matrix of the spin. From this we calculate the representation of the dynamical map given by Eq.\,(\ref{eq:RhoReducedSystemPropagator}). Using Eq.\,(\ref{eq:BlochVectorTimeEvolution}), the two quantities $\mathbf{M}(t)$ and $\vec{b}(t)$ can be calculated numerically.

We focus on the zero temperature limit, where it is known that the spin localizes in its initial state for an Ohmic spectral density in the strong coupling regime \cite{Bray1982,Chakravarty1982,Wang2019}. Consequently, the harmonic oscillators of the bath are initially all in their ground state. The spin is initially in a pure state of the form
\begin{align}
\ket{\psi(0)}_{\rm spin} &= \cos \frac{\theta}{2} \ket{\uparrow} + {\rm e}^{i \varphi} \sin \frac{\theta}{2} \ket{\downarrow},
\end{align}
where $\theta$ and $\varphi$ parameterize the direction of the spin at time $t=0$.
\subsection{Ohmic Spectral Density}
We begin our discussion with the Ohmic spectral density. It is known that in the scaling limit ($\omega_c/\Delta \to \infty$), the dynamics of the spin can be grouped into three qualitatively different regimes, comprising coherent decay for weak system-environment coupling ($\alpha<0.5$), incoherent decay (intermediate coupling, $0.5 < \alpha <1)$, and localization (strong coupling $\alpha>1$) \cite{Weiss1999, Wang2008, Leggett1987}. Thus, we expect that for $\alpha<1$ the spin relaxes to a unique asymptotic state, whereas for $\alpha>1$ the asymptotic state depends on the initial state. It is also known that for finite $\omega_c/\Delta$ both critical couplings shift to larger values \cite{Wang2019,Thoss2001,Wang2008,Wang2010}.

\begin{figure}[t]
\includegraphics[scale=0.5]{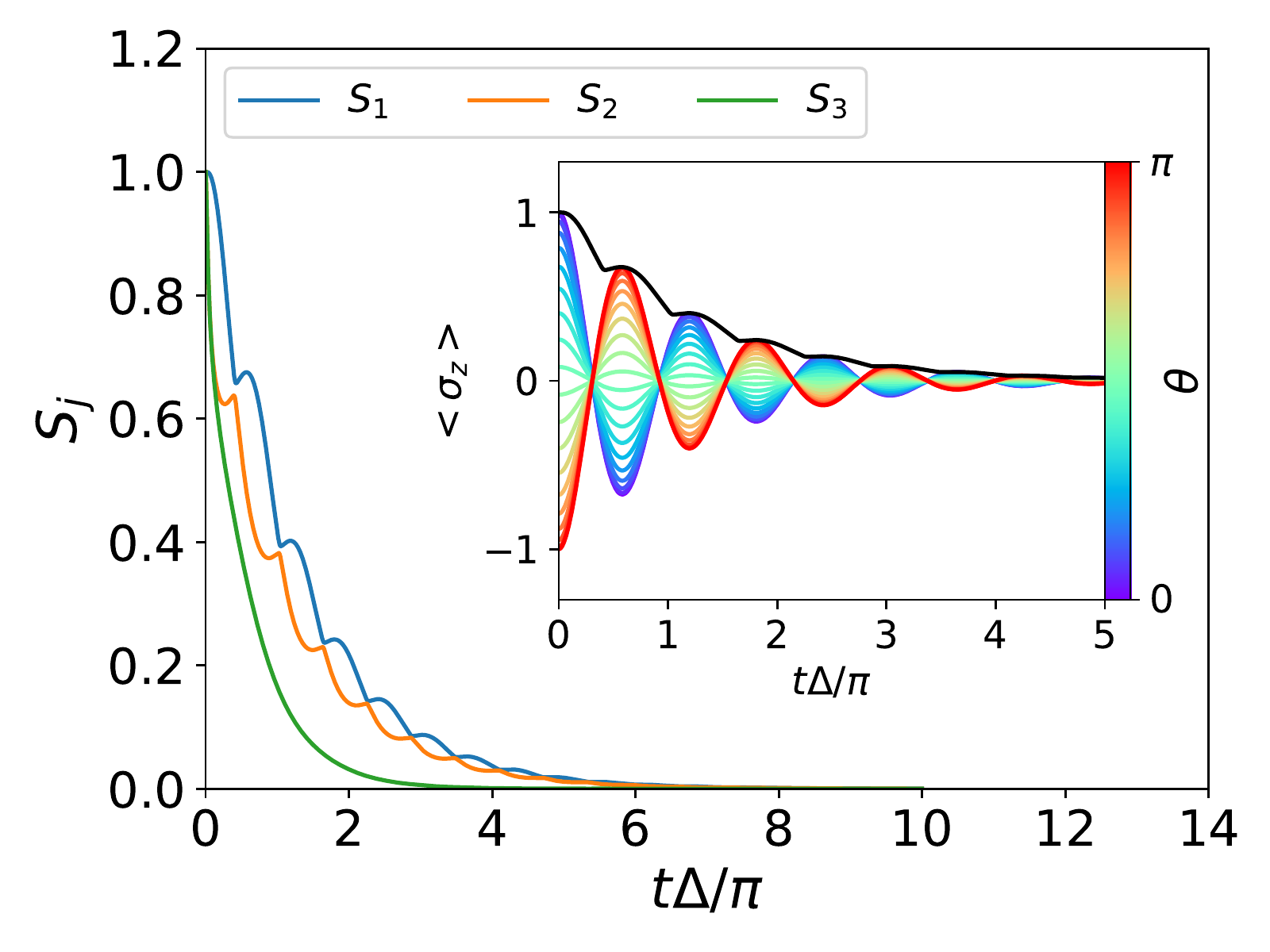}
\caption{Singular values $S_j(t)$ of ${\bf M}(t)$ as a function of time for the Ohmic spectral density for $\omega_c=20\Delta$ and $\alpha=0.1$. The inset shows $\braket{\sigma_z}(t)$ for different initial states of the spin parametrized by $\theta$. For all initial states $\varphi=0$. The black line in the inset shows $S_{\textrm{max}}(t)$.}\label{fig:unique_state_ohmic_density}
\end{figure}

In Fig.\,\ref{fig:unique_state_ohmic_density}, the singular values of ${\bf M}(t)$ are shown for a weak coupling $\alpha$ and large $\omega_c$. It directly follows from Eq.\,(\ref{eq:BlochVectorTimeEvolution}) that ${\bf M}(0)=\mathds{1}$, and thus, all three singular values of ${\bf M}(0)$ are initially one. Over time, they decay to zero, reflecting the vanishing influence of the initial state on the dynamics. The first two singular values exhibit periodic modulations. We find that $S_1(t)>S_2(t)$ for all times, i.e. the two singular values never cross. The third singular value decays monotonically. 

In the weak coupling and large $\omega_c$ limit an approximate analytic solution can be used to connect the behavior of the singular values with the dynamics of the spin \cite{Leggett1987,Weiss1999,Thoss2001}. The equations for the Bloch vector $\vec{a}(t)$, as well as the derivation of ${\bf M}(t)$ and $\vec{b}(t)$ for the weak coupling limit are provided in Appendix \ref{appendix:C}. The analytic solution reveals that the periodically modulated singular values $S_1(t)$ and $S_2(t)$ are related to the coherent decay of $\braket{\sigma_y}(t)$ and $\braket{\sigma_z}(t)$. The two singular values are exponentially damped, with the same damping as $\braket{\sigma_y}(t)$ and $\braket{\sigma_z}(t)$. The frequency of the periodic modulations of $S_1(t)$ and $S_2(t)$ is given by $2\tilde{\Delta}$, where $\tilde{\Delta}$ denotes the renormalized frequency of the spin. The monotonically decaying singular value $S_3(t)$ describes the monotonic decay of $\braket{\sigma_x}(t)$. Both, the singular value $S_3(t)$ and $\braket{\sigma_x}(t)$, decay exponentially with the same decay rate.

In the inset of Fig.\,\ref{fig:unique_state_ohmic_density}, the vanishing influence of the initial state for long times is exemplified for the expectation value of $\sigma_z$. As predicted by the vanishing of the singular values, $\braket{\sigma_z}(t)$ relaxes to an equilibrium value as $t \to \infty$ independent of its initial value. The black line is the largest singular value which provides a bound for the influence of the initial state at all times. Note that according to Eq.\,(\ref{eq:BoundExpectationValues}) the quantity $\delta_{1,2}(t;\sigma_z)$ is bounded by $4 S_{\textrm{max}}(t)$, while we find $\delta_{1,2}(t;\sigma_z)<2 S_{\textrm{max}}(t)$. This is related to the fact that in the derivation of Eq.\,(\ref{eq:BoundExpectationValues}) no property of the observable $O$ was used. For $O=\sigma_z$ one can employ that the expectation value of $\sigma_x$ and $\sigma_y$ vanishes in both eigenstates of $\sigma_z$. Using this, one can show that $\delta_{1,2}(t;\sigma_z)$ is bounded by $2 S_{\textrm{max}}(t)$.

\begin{figure}[b]
\includegraphics[scale=0.5]{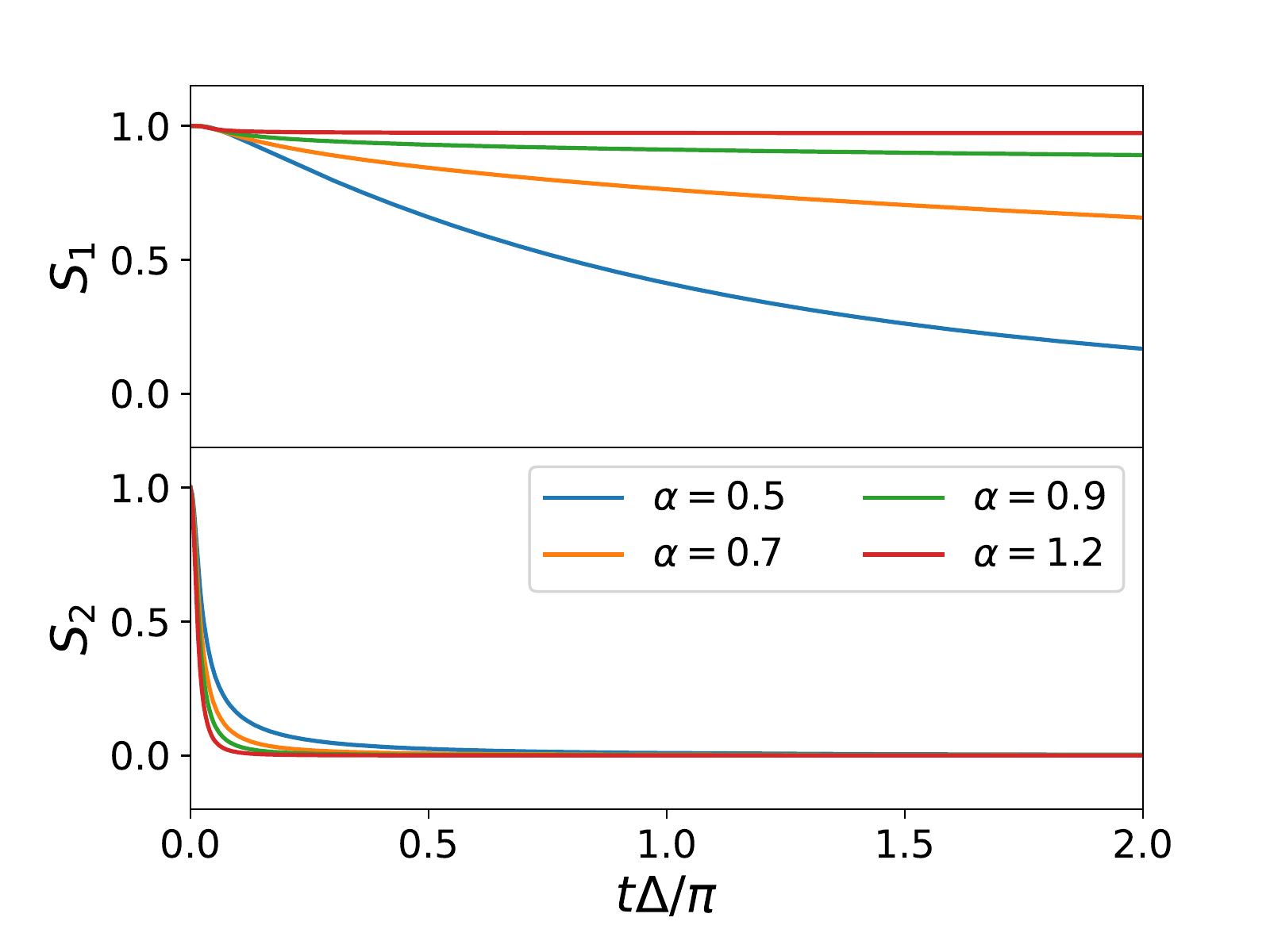}
\caption{The first and second singular value of ${\bf M}(t)$ as a function of time for different values of $\alpha$ and $\omega_c=20\Delta$.}\label{fig:svd_coupling_ohmic_density}
\end{figure}

For moderate couplings, $0.5 < \alpha <1$, the spin exhibits incoherent decay. In this regime the decay slows down as the coupling strength is increased until the spin eventually localizes at $\alpha\approx 1$. In Fig.\,\ref{fig:svd_coupling_ohmic_density}, the first and second singular value of ${\bf M}(t)$ are shown for $\omega_c=20 \Delta$ for different values of the coupling strength $\alpha$. The third singular value shows a very similar behavior as the second one and is thus not shown. Similar to the weak coupling regime, we find that all singular values of ${\bf M}(0)$ are initially one. For $0.5 < \alpha <1$, the three singular values exhibit a monotonic decay to zero. They show, however, an opposite trend upon increasing the coupling strength. $S_2(t)$ and $S_3(t)$ decay faster for increasing coupling strength, while the decay of $S_1(t)$ slows down and eventually approaches a non-zero value as $t \to \infty$ for $\alpha \gtrsim 1$. This is consistent with the transition from incoherent decay to localization as the coupling strength approaches $\alpha \approx 1$. One way to understand this localization was introduced by Silbey and Harris by means of a renormalized system frequency $\tilde{\Delta}$ \cite{Silbey1984,Harris1985}. For the Ohmic spectral density they used a variational Polaron transformation to calculate $\tilde{\Delta}$, showing that in the limit $\Delta/\omega_c \to 0$ the renormalized system frequency vanishes above a critical coupling $\alpha_c$, i.e. $\tilde{\Delta}=0$ for $\alpha> \alpha_c$. Due to the vanishing of the effective coupling, the spin is frozen in its initial state explaining the dependence of the asymptotic state on the initial state. 

\begin{figure}[t]
\includegraphics[scale=0.5]{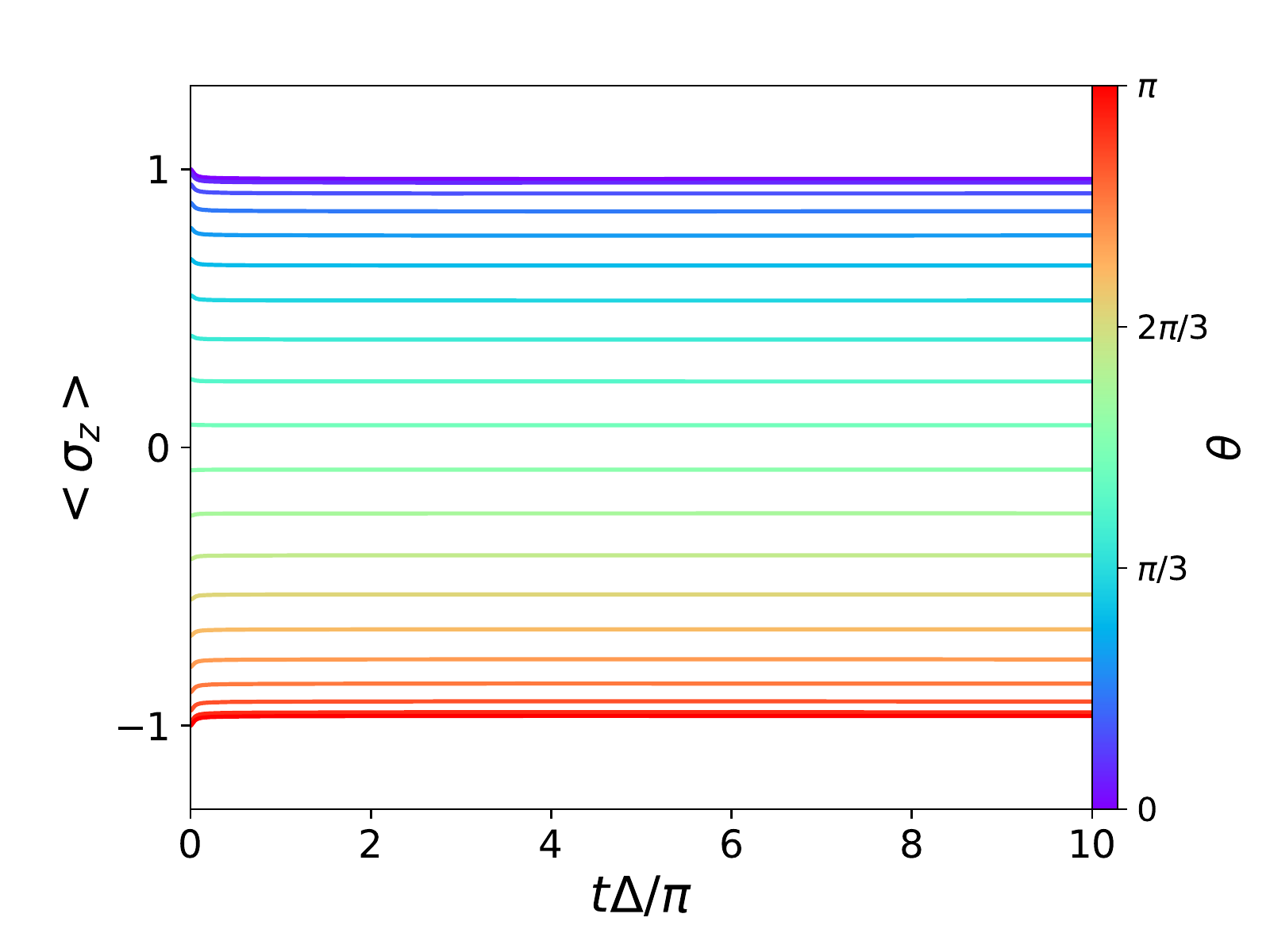}
\caption{$\braket{\sigma_z}$ as a function of time for different initial states parametrized by $\theta$. In this plot $\omega_c=20 \Delta$, $\alpha=1.2$ and $\varphi=0$. The results show that the asymptotic state of the spin depends on the initial state of the spin.}\label{fig:non_unique_state_ohmic_density}
\end{figure}

The dependence of the asymptotic state on the initial state for strong coupling ($\alpha=1.2$) is exemplified in Fig.\,\ref{fig:non_unique_state_ohmic_density} for the expectation value of $\sigma_z$. For all initial states of the spin, the expectation value becomes stationary. However, the stationary value of $\braket{\sigma_z}(t)$ depends on the initial state of the spin. As discussed above, for $ \alpha\gtrsim 1$, one singular value of ${\bf M}(t)$ approaches a non-zero value as $t \to \infty$, implying that the image of ${\bf M}_\infty$ is one dimensional. Thus, ${\bf M}_\infty$ projects the set of initial states, the Bloch sphere, onto a one dimensional subset. This also implies that some initial states are mapped to the same asymptotic state. In order to classify those initial states which are mapped to the same asymptotic state, we consider the singular value decomposition of $\mathbf{M}_{\infty}$, in the following denoted by $\mathbf{V}_{\infty} \mathbf{S}_{\infty} \mathbf{W}_{\infty}^T$. Assuming that only one singular value is non-vanishing, the asymptotic state of the spin can be written as
\begin{align}
{\bf a}_{\infty} = s_{\infty,1} \braket{ {\bf w}_{\infty, 1},  {\bf a}(0)} {\bf v}_{\infty, 1} + {\bf b}_{\infty, 1},
\end{align}
where $\mathbf{v}_{\infty, 1}$ ($\mathbf{w}_{\infty, 1}$) are the vectors formed by the first column of the matrix $\mathbf{V}_\infty$ ($\mathbf{W}_\infty$), respectively, and $\braket{x,y}=\sum_{n} x_n y_n$ is the standard scalar product between real vector $\mathbf{x}$ and $\mathbf{y}$. All initial states $\mathbf{a}(0)$, for which the scalar product $\braket{\mathbf{w}_{\infty,1},\mathbf{a}(0)}$ is equal are mapped to the same asymptotic state. This means that all initial states which are in a plane orthogonal to $\mathbf{w}_{\infty, 1}$ are mapped to the same asymptotic state. For the parameters considered in Fig.\,\ref{fig:non_unique_state_ohmic_density}, i.e. $\omega_c=20\Delta$ and $\alpha=1.2$, we find $\mathbf{w}_{\infty,1}^T \approx (0, ~ 0.13, ~ 0.99)$.

\begin{figure}[b]
\vspace*{0.5cm}
\includegraphics[scale=0.46]{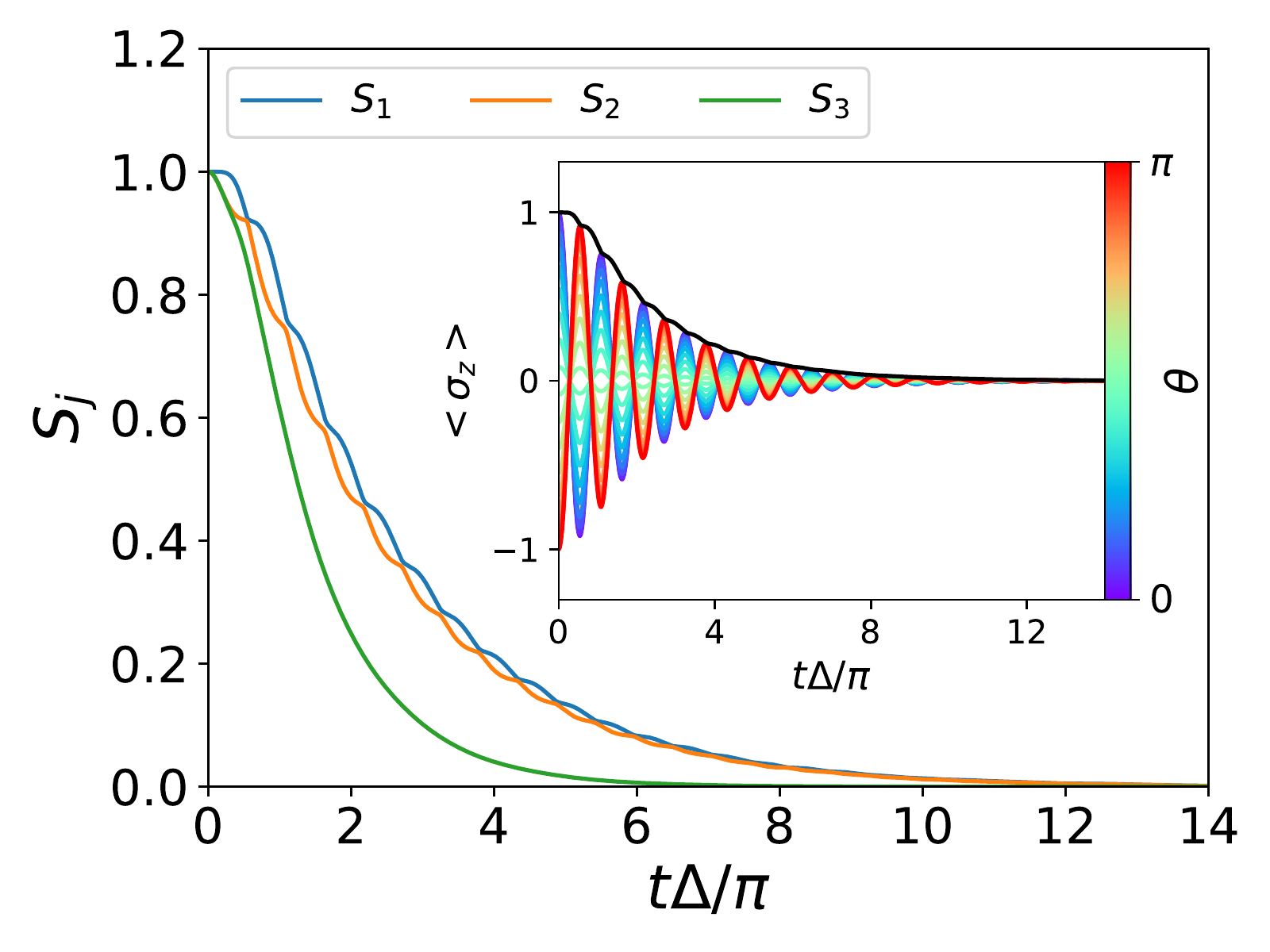}
\caption{Singular values $S_j$ of ${\bf M}(t)$ as a function of time for the gapped spectral density for $\alpha=0.1$. The inset shows $\braket{\sigma_z}(t)$ for different initial states of the spin, where $\varphi=0$. The black line in the inset shows the largest singular value of ${\bf M}(t)$.} \label{fig:unique_state_gapped_density}
\end{figure}

\subsection{Gapped Spectral Density}
To demonstrate the behavior of ${\bf M}(t)$ for a case where the open system does not approach an asymptotic state, we consider the gapped spectral density defined in Eq.\,(\ref{eq:SpectralDensityGapped}). 

Fig.\,\ref{fig:unique_state_gapped_density} shows the three singular values of ${\bf M}(t)$ for weak coupling, $\alpha=0.1$. The qualitative behavior is very similar to the Ohmic spectral density, i.e. all singular values show an overall decay to zero. $S_1(t)$ and $S_2(t)$ exhibit a periodically modulated decay, where the frequency of the periodic modulations is twice the frequency of the spin. $S_3(t)$ decays in a non-oscillatory way. The vanishing influence of the initial state is exemplified in the inset of Fig.\,\ref{fig:unique_state_gapped_density} for the expectation value of $\sigma_z$. The expectation value of $\sigma_z$ decays to zero for all values of $\theta$. As indicated by the black line the quantity $\delta_{1,2}(t;\sigma_z)$, is again bounded by $2S_{\textrm{max}}(t)$.

\begin{figure}[t]
\includegraphics[scale=0.5]{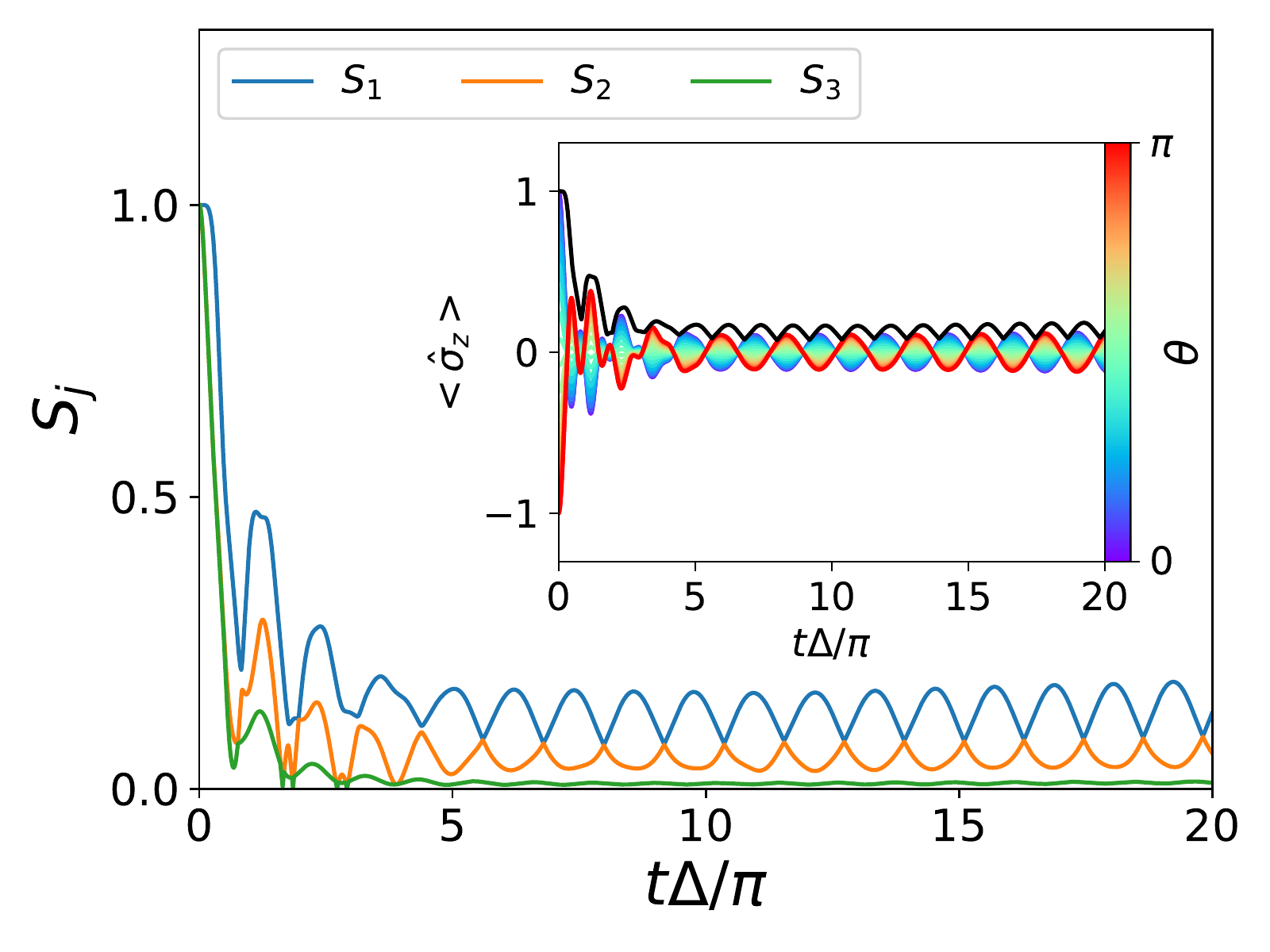}
\caption{Singular values $S_j$ of $M(t)$ as a function of time for $\alpha=1.2$. The inset shows the expectation value of $\sigma_z$ for different initial states of the open system, where $\varphi=0$. The black line in the inset shows the largest singular value of $M(t)$.}\label{fig:non_unique_state_gapped_density}
\end{figure}

In the strong coupling regime, however, we find that the spin does not approach an asymptotic state. This is illustrated in Fig.\,\ref{fig:non_unique_state_gapped_density}, which shows the three singular values of ${\bf M}(t)$ for $\alpha=1.2$.
All three singular values exhibit a fast initial decay. Unlike in the regimes discussed so far, the singular values cross for $t\Delta/\pi \lesssim 5$. After the initial decay, the smallest singular value decays further and remains close to zero, while the other two singular values of ${\bf M}(t)$ exhibit undamped oscillations, which indicates the non-existence of an asymptotic state, i.e. the state of the open system remains time-dependent at all times. For these longer times, we find that $S_1(t)>S_2(t)$ for all times, i.e. the two singular values do not cross. Similar to the above discussed cases, the period of the oscillations of the singular values have twice the frequency of the spin. In the inset of Fig.\,\ref{fig:non_unique_state_gapped_density}, the expectation value of $\sigma_z$ is shown demonstrating that the expecation value does not approach a stationary state and that the initial state has an influence on $\braket{\sigma_z}(t)$ at all times. The influence of the initial state on $\braket{\sigma_z}(t)$ is again bounded by $2S_{\textrm{max}}(t)$ at all times.

This behavior can be understood in a similar way as the localization in the Ohmic case. Using the same variational polaron transformation as Silbey and Harris \cite{Silbey1984, Harris1985}, one can show numerically that the renormalized system frequency $\tilde{\Delta}$ for the gapped spectral density is a monotonically decreasing function of the coupling strength. For sufficiently strong coupling $\alpha$ the renormalized system frequency is below the smallest bath frequency. In this case, the bath cannot act as a true environment and fails to equilibrate the spin. For $\alpha=1.2$, we find that $\tilde{\Delta}$ is close to the smallest bath frequency, resulting in a partial decoherence of the spin on short time scales. However, the environment cannot equilibrate the spin completely and the state of the spin remains time-dependent at all times.

\section{Conclusion}
In this paper we have proposed a measure to quantify the influence of the initial state of an open system on its dynamics, based on the dynamical map, a quantity which describes the time evolution of an open system in the presence of an environment. Using this measure it is possible to investigate the asymptotic behavior of a quantum system in contact with an environment, and to quantify the information stored in local observables. We have demonstrated our theoretical concepts for the well-known spin-boson model and identified three qualitatively different long-time behaviors, which can be distinguished by considering the singular values of the dynamical map. We note that it is possible to reconstruct the measure from local expectation values alone, making it experimentally accessible. It might be of interest to apply our measure to analyze numerically observed bistabilities in more complex systems, like the one discussed in Ref.\,[\onlinecite{Wilner2013}]. The investigation of local memory in many-body quantum systems, e.g. systems exhibiting many-body localization, based on the dependence of the initial state of a subsystem is another interesting subject of future work.


\section{Acknowledgments}
We thank Haobin Wang for providing the ML-MCTDH
code used in this work. We also thank Tobias Schätz and Ulrich Warring for discussions on the experimental realization of the spin-boson model and for providing the experimental parameters used to fit the gapped spectral density. This work was supported by the German Research Foundation (DFG) through FOR 5099. Furthermore, support by the state of Baden-Württemberg through bwHPC and the DFG through Grant No. INST 40/575-1 FUGG (JUSTUS 2 cluster) is gratefully acknowledged.

\clearpage

\clearpage
\appendix
\section{Proof of the bound for $\delta_{1,2}(t; O)$}
\label{appendix:A}
In the following we proof the bound for $\delta_{1,2}(t;O)$ defined as
\begin{align}
\delta_{1,2}(t;O) &= |{\rm tr}\{O(\rho^1(t)-\rho^2(t)) \}|,
\end{align}
where $O$ is some observable. The difference between the two density matrices $\rho^1(t)$ and $\rho^2(t)$ can be written as
\begin{align}
\rho^1(t)-\rho^2(t) &= \frac{1}{2} \sum_{n=1}^{N^2-1} \big(a_n^1(t) - a_n^2(t)\big)T_n
\end{align}
We evaluate the trace in the eigenbasis of $O$ yielding
\begin{align}
\delta_{1,2}(t;O) &= \frac{1}{2} \big| \sum_{n=1}^{N^2-1} \sum_{a=1}^{N}  \big(a^1_n(t) -a^2_n(t)\big)  o_a \braket{o_a| T_n| o_a} \big|.
\end{align}
Using the triangular inequality and rearranging the sums yields
\begin{align}
\delta_{1,2}(t;O) &\leq  \frac{1}{2} \sum_{n=1}^{N^2-1} \big| (a^1_n(t) -a^2_n(t)) \big| \sum_{a=1}^{N} |o_a| ~ \big| \braket{o_a| T_n| o_a} \big|
\end{align}
Let $o_{\rm max}$ be the eigenvalue of $O$ with the largest absolute value. The second sum is bounded by
\begin{align}
\sum_{a=1}^{N} |o_a| ~ \big| \braket{o_a| T_n| o_a} \big| \leq |o_{\rm max}| \sum_{a=1}^{N} \big| \braket{o_a| T_n| o_a} \big| 
\end{align}
Similarly, it can be shown that the expectation value of $T_n$ in an eigenstate of $O$ is bounded by the eigenvalue of $T_n$ with the largest eigenvalue, which in this case depends on $n$:
\begin{align}
\big| \braket{o_a| T_n| o_a} \big| \leq t_{n,{\rm max}}
\end{align}
From the definition of the generators of $SU(N)$, see [\onlinecite{Kimura2003}], it directly follows that $t_{n, {\rm max}}$ is bounded by $\sqrt{2}$. Using this, one obtains a bound for $\delta_{1,2}(t;O)$ reading
\begin{align}
\delta_{1,2}(t;O) &\leq  \frac{1}{2} |o_{\rm max}| \sqrt{2} N ~  \sum_{n=1}^{N^2-1} | a^1_n(t) -a^2_n(t) |
\end{align}
where the factor $N$ originates from the sum over the eigenbasis of $O$. The remaining sum is the $L^1$ norm of the vector $\vec{a}^1(t) - \vec{a}^2(t)$, i.e.
\begin{align}
\delta_{1,2}(t;O) &\leq  \frac{1}{2} |o_{\rm max}| \sqrt{2} N ~ ||  \vec{a}^1_n(t) - \vec{a}^2_n(t) ||_1.
\end{align}
To relate the distance between the time-dependent Bloch vectors to the initial distance we first employ that the $L^1$ norm of a vector $\vec{x}$ is bounded by the Euclidean norm as $||\vec{x}||_1 \leq \sqrt{N} || \vec{x}||_2$. Using the singular value decomposition of ${\bf M}(t)$ one can write
\begin{align}
||  \vec{a}^1_n(t) - \vec{a}^2_n(t) ||_2 = || {\bf V}(t) {\bf S}(t) {\bf W}^T(t) \big(\vec{a}^1_n(0) - \vec{a}^2_n(0)\big) ||_2.
\end{align}
Using the fact that ${\bf S}$ is a positive semidefinite diagonal matrix, it is easy to show that $|| {\bf S}\vec{x}||_2 < s_{\rm max} || \vec{x}||_2$, where $s_{\rm max}$ is the largest diagonal element of ${\bf S}$. Using this, together with the fact that the Euclidean norm is invariant under orthogonal transformations one can write
\begin{align}
\delta_{1,2}(t;O) &\leq \frac{N^{3/2}}{\sqrt{2}} |o_{\rm max}| ~  s_{\rm max}(t)~ ||\vec{a}^1(0)-\vec{a}^2(0)||_2,
\end{align}
which finishes the proof of Eq.\,(\ref{eq:BoundExpectationValues}).
\section{Gapped spectral density}
\label{appendix:B}
The gapped spectral density originates from an experimental realization of the spin-boson model using trapped ions \cite{Clos2016}. Using a linear Paul trap a spin-boson model with up to five bosonic degrees of freedom was realized. The distinct feature of the experiment is that the smallest frequency of the bosonic environment is determined by the frequency of the trap, and is thus always larger than zero. We consider the continuum limit of the experimental spectral density which is derived as follows. The spectral density of the five ion model is given by 
\begin{align}
J_5(\omega) &= \frac{\pi}{4} \alpha \sum_{n=1}^5 \omega_n \mathcal{M}_n^2 \delta(\omega - \omega_n),
\end{align}
where $\omega_n$ and $\mathcal{M}_n$ are determined from experimental parameters and the geometry of the trap \cite{Clos2016,Porras2004, Porras2008}. $\alpha$ denotes the coupling strength between the spin and the environment. To study the dynamics for different coupling strengths $\alpha$, we fit the experimental parameters for $J_5(\omega)/\alpha$ with the function 
\begin{align}
J_{\rm fit}(\omega) &=  \frac{\pi}{4} a (\omega-b) {\rm e}^{-\big(\frac{\omega-b}{c}\big)^3},
\end{align}
where $a$, $b$ and $c$ are fitting parameters. The gapped spectral density we consider is thus given by 
\begin{align}
J_G(\omega/\omega_1)&= \alpha \frac{\pi}{4} a(\omega-b) {\rm e}^{-\big(\frac{\omega-b}{c}\big)^3},
\end{align}
where $\alpha$ denotes the coupling strength, and $(a, b, c)=(0.677, 0.541, 1.280)$. 

\section{Weak coupling solution}
\label{appendix:C}
In the weak coupling limit a perturbative approach can be used to analyze the connection between the dynamics of the spin and the singular values of ${\bf M}(t)$ analytically. The starting point is the second-order time convolutionless (TCL2) master equation \cite{Shibata1977,Chaturvedi1979,Shibata1980}. Within the TCL2 approach, the equations of motion for the three expectation values read \cite{Breuer2007}
\begin{align}
\partial_t \braket{\sigma_x}(t) &= - \Gamma_{xx}(t) \braket{\sigma_x}(t)  -\Gamma_x(t) \label{appC:TdpDglX} \\
\partial_t \braket{\sigma_y}(t) &= - 2\Delta \braket{\sigma_z}(t) - \Gamma_{yz}(t) \braket{\sigma_z}(t) - \Gamma_{yy}(t) \braket{\sigma_y}(t) \label{appC:TdpDglY} \\
\partial_t \braket{\sigma_z}(t) &= 2 \Delta \braket{\sigma_y}(t) \label{appC:TdpDglZ},
\end{align}
where the time-dependent rates $\Gamma_{ij}(t)$ are determined by the spectral density $J(\omega)$ and $\Delta$, and are defined in [\onlinecite{Breuer2007}]. To solve the equations (\ref{appC:TdpDglX}), (\ref{appC:TdpDglY}), and (\ref{appC:TdpDglZ}) analytically, we consider the stationary rate approximation, i.e. we replace the time-dependent rates with their long time limit, in the following denoted with $ \Gamma_{ij}  \coloneqq \lim_{t \to \infty} \Gamma_{ij}(t) $. The resulting equations of motion read
\begin{align}
\partial_t \braket{\sigma_x}(t) &= - \Gamma_{xx} \braket{\sigma_x}(t)  - \Gamma_x \label{appC:StatDglX}\\
\partial_t \braket{\sigma_y}(t) &= - 2\Delta \braket{\sigma_z}(t) - \Gamma_{yz} \braket{\sigma_z}(t) - \Gamma_{yy} \braket{\sigma_y}(t)\label{appC:StatDglY} \\
\partial_t z(t) &= 2 \Delta \braket{\sigma_y}(t)\label{appC:StatDglZ}.
\end{align}
Equations (\ref{appC:StatDglX}), (\ref{appC:StatDglY}), and (\ref{appC:StatDglZ}) constitute a system of first order autonomous differential equations, and thus, can be solved analytically. Their solution read
\begin{align}
\braket{\sigma_x}(t) =  &{\rm e}^{-\Gamma_{xx} t } \braket{\sigma_x}(0) - \frac{\Gamma_{x}}{\Gamma_{xx}} (1-{\rm e}^{-\Gamma_{xx} t})  \label{appC:SolX} \\
\braket{\sigma_y}(t) = &{\rm e}^{-\Gamma_{yy}/2 t} \bigg(  \cos(\tilde{\Delta} t) \braket{\sigma_y}(0)  -\frac{\Gamma_{yy}}{2 \tilde{\Delta}}\sin(\tilde{\Delta} t) \braket{\sigma_y}(0)  \nonumber \\
&\qquad -  \frac{2\Delta - \Gamma_{yz}}{\tilde{\Delta}}  \sin(\tilde{\Delta} t) \braket{\sigma_z}(0) \bigg) \label{appC:SolY} \\
\braket{\sigma_z}(t) = &{\rm e}^{- \Gamma_{yy}/2 t} \bigg( \frac{2\Delta}{\tilde{\Delta}} \sin(\tilde{\Delta} t) \braket{\sigma_y}(0)+ \cos(\tilde{\Delta} t) \braket{\sigma_z}(0)   \nonumber \\
&+ \frac{\Gamma_{yy}}{2 \tilde{\Delta}} \sin(\tilde{\Delta} t) \braket{\sigma_z}(0)  \bigg)\label{appC:SolZ},
\end{align}
where $\tilde{\Delta}$ denotes the renormalized frequency of the spin and is given by $\tilde{\Delta} = \frac{1}{2}\sqrt{8 \Delta(2\Delta - \Gamma_{yz}) - \Gamma_{yy}^2}$. From equations (\ref{appC:SolX}), (\ref{appC:SolY}), and (\ref{appC:SolX}) the quantities $\vec{b}(t)$ and ${\bf M}(t)$ can be identified as
\begin{widetext}
\begin{align}
\vec{b}(t) &= 
\begin{pmatrix}
\frac{\Gamma_x}{\Gamma_{xx}} (1-{\rm e}^{-\Gamma_{xx} t}) & 0 & 0
\end{pmatrix}
^T \\
{\bf M}(t) &=
\begin{pmatrix}
{\rm e}^{-\Gamma_{xx} t} & 0 & 0 \\
0 & {\rm e}^{- \Gamma_{yy}/2 t}\big( \cos(\tilde{\Delta} t) - \frac{\Gamma_{yy}}{2 \tilde{\Delta}}\sin(\tilde{\Delta} t) \big) & -\frac{2 \Delta- \Gamma_{yz}}{\tilde{\Delta}}{\rm e}^{- \Gamma_{yy}/2 t} \sin(\tilde{\Delta} t) \\
 0 &  \frac{2 \Delta}{\tilde{\Delta}} {\rm e}^{- \Gamma_{yy}/2 t}\sin(\tilde{\Delta} t) & {\rm e}^{- \Gamma_{yy}/2 t} \big( \cos(\tilde{\Delta} t) + \frac{\Gamma_{yy}}{2 \tilde{\Delta}}\sin(\tilde{\Delta} t) \big)
\end{pmatrix}.
\end{align}
\end{widetext}
Note that in this perturbative treatment, the time-evolution of $\braket{\sigma_y}(t)$ and $\braket{\sigma_z}(t)$ is independent of $\braket{\sigma_x}(t)$, and thus ${\bf M}(t)$ is block diagonal.  The singular values of ${\bf M}(t)$ are given by the square root of the eigenvalues of ${\bf M}^T(t) {\bf M}(t)$. Thus, the singular value which is associated to the dynamics of $\braket{\sigma_x}$ is given by ${\rm e}^{-\Gamma_{xx} t}$. The other two singular values are obtained by diagonalizing the $2 \times 2$ block associated to the dynamics of $\braket{\sigma_y}$ and $\braket{\sigma_z}$. The two singular values $S_{\pm}(t)$ are given by
\begin{align}
S_\pm(t) &= \frac{{\rm e}^{-\Gamma_{yy}/2 t}}{2} \bigg( A(t) \pm \sqrt{A^2(t) - 4~B(t) } \bigg),
\end{align}
where we defined
\begin{align}
A(t) &= \bigg(\frac{\Gamma_{yy}^2}{2\tilde{\Delta}^2}+\frac{4 \Delta^2}{\tilde{\Delta}^2}-\frac{(2\Delta-\Gamma_{yz})^2}{\tilde{\Delta}^2}\bigg)\sin^2(\tilde{\Delta}t)\nonumber \\
&~~~~~+2 \cos^2(\tilde{\Delta} t) \nonumber \\
B(t) &= \frac{\Gamma_{yy}^4}{8\tilde{\Delta}^4}\sin^4(\tilde{\Delta}t) - 2\frac{\Gamma_{yy}^2}{4\tilde{\Delta}^2} \sin^2(\tilde{\Delta}t)\cos^2(\tilde{\Delta}t)\nonumber \\
&\quad -\frac{4 \Delta^2}{\tilde{\Delta}^2} \frac{(2\Delta-\Gamma_{yz})^2}{\tilde{\Delta}^2}\sin^4(\tilde{\Delta}t) + \cos^4(\tilde{\Delta}t) \nonumber
\end{align}
Since $A(t)$ and $B(t)$ are periodic with period $2\tilde{\Delta}$, it follows that $S_\pm(t)$ are described by damped oscillations, where the damping is given by ${\rm e}^{-\Gamma_{yy}/2 t}$ and the period of the oscillation is $2 \tilde{\Delta}$. These two singular values describe the coherent decay of $\braket{\sigma_y}(t)$ and $\braket{\sigma_z}(t)$.

\end{document}